\documentclass[sigchi, nonacm]{acmart}

\AtBeginDocument{%
  }

\usepackage{tikz}
\usepackage{amsmath}
\usepackage{csquotes}
\usepackage{multirow}
\usepackage{adjustbox}
\usepackage{xspace}
\usepackage{cleveref}
\usepackage[utf8]{inputenc}
\usepackage{enumitem}
\usepackage{hyperref}
\usepackage{csquotes}

\usepackage{lipsum}

\newif\ifdraft
\drafttrue

\definecolor{beaublue}{rgb}{0.74, 0.83, 0.9}
\ifdraft 
    \newcommand{\TODOS}[1]{{\color{purple} \textbf{TODO:} #1}}

\else
    \newcommand{\TODOS}[1]{}

\fi

\newcommand{\tap}{\texttt{TAP}\xspace}
\newcommand{\sign}{\texttt{SIGN}\xspace}
\newcommand{\pink}{\texttt{PIN-K}\xspace}
\newcommand{\pinf}{\texttt{PIN-F}\xspace}

\usepackage{titlesec}

\begin{document}

\title{How Interactions Influence Users' Security Perception of Virtual Reality Authentication?}

\author{Jingjie Li}
\authornote{The work was done while the authors were at Visa Research.}
\affiliation{%
  \institution{University of Wisconsin--Madison}
  \city{Madison, Wisconsin}
  \country{USA}}
\email{jingjie.li@wisc.edu}

\author{Sunpreet Singh Arora}
\affiliation{%
  \institution{Visa Research}
  \city{Palo Alto, California}
  \country{USA}}
\email{sunarora@visa.com}

\author{Kassem Fawaz}
\affiliation{%
  \institution{University of Wisconsin--Madison}
  \city{Madison, Wisconsin}
  \country{USA}}
\email{kfawaz@wisc.edu}

\author{Younghyun Kim}
\affiliation{%
  \institution{University of Wisconsin--Madison}
  \city{Madison, Wisconsin}
  \country{USA}}
\email{younghyun.kim@wisc.edu}

\author{Can Liu}
\affiliation{%
  \institution{Visa Research}
  \city{Austin, Texas}
  \country{USA}}
\email{caliu@visa.com}

\author{Sebastian Meiser}
\authornotemark[1]
\affiliation{%
  \institution{University of Lübeck}
  \city{Lübeck}
  \country{Germany}}
\email{sebastian@meiser-web.de}

\author{Mohsen Minaei}
\affiliation{%
  \institution{Visa Research}
  \city{Palo Alto, California}
  \country{USA}}
\email{mominaei@visa.com}

\author{Maliheh Shirvanian}
\authornotemark[1]
\affiliation{%
  \institution{Netflix Inc.}
  \city{Los Gatos, California}
  \country{USA}}
\email{maliheh21@gmail.com}

\author{Kim Wagner}
\affiliation{%
  \institution{Visa Research}
  \city{Palo Alto, California}
  \country{USA}}
\email{kwagner@visa.com}

\begin{abstract}

Users readily embrace the rapid advancements in virtual reality (VR) technology within various everyday contexts, such as gaming, social interactions, shopping, and commerce. In order to facilitate transactions and payments, VR systems require access to sensitive user data and assets, which consequently necessitates user authentication. However, there exists a limited understanding regarding how users' unique experiences in VR contribute to their perception of security. In our study, we adopt a research approach known as ``technology probe'' to investigate this question. Specifically, we have designed probes that explore the authentication process in VR, aiming to elicit responses from participants from multiple perspectives. These probes were seamlessly integrated into the routine payment system of a VR game, thereby establishing an organic study environment. Through qualitative analysis, we uncover the interplay between participants' interaction experiences and their security perception. Remarkably, despite encountering unique challenges in usability during VR interactions, our participants found the intuitive virtualized authentication process beneficial and thoroughly enjoyed the immersive nature of VR. Furthermore, we observe how these interaction experiences influence participants' ability to transfer their pre-existing understanding of authentication into VR, resulting in a discrepancy in perceived security. Moreover, we identify users' conflicting expectations, encompassing their desire for an enjoyable VR experience alongside the assurance of secure VR authentication. Building upon our findings, we propose recommendations aimed at addressing these expectations and alleviating potential conflicts.

\end{abstract}

\maketitle
\pagestyle{plain}

\section{Introduction}

Virtual reality (VR) systems immerse individuals in a digital world, one that simulates real-world interactions with objects and characters~\cite{burdea-vrt03}.
In addition to specialized use cases (e.g., military training and healthcare~\cite{rizzo-jcpms11}), VR technology is seeing widespread adoption in everyday settings, such as gaming, social interactions, shopping, and commerce~\cite{freeman-cscw21,speicher-imwut17,hock-chi17}. Payment features in VR empowers these activities and contributes to the growth of VR economics~\cite{lau_2022}.

To enable payments, VR systems access sensitive user data and assets, raising the need to authenticate users. VR service providers deploy user authentication methods borrowed from traditional platforms to verify users' identities, such as using passwords and personal identification numbers (PINs). However, VR presents a unique context where users interact with digital objects to perform routine activities, such as payment, which were once limited to conventional platforms. Recent research has shown that the context in which authentication is used (e.g., where and for what purpose) affects how users perceive the security of authentication. For example, users feel insecure when using an ATM in a crowded space~\cite{mathis-vr22}. There is a critical need to understand \textbf{how the unique experience in VR contributes to users' security perception of payment authentication.} With such understanding, we can guide the future design of authentication methods that are both secure and usable in the growing VR commerce ecosystem. Our paper provides this understanding by investigating these related research questions.

\begin{itemize}[noitemsep]

\item \textbf{RQ1 -- Interaction Experiences:} What are the factors in users' interaction experiences that contribute to their security perception of authentication in VR?
\item \textbf{RQ2 -- Influences on Security Perception:} How do users' interaction experiences influence their security perceptions of authentication in VR?
\item \textbf{RQ3 -- Understanding User Expectations:} What are the tensions in users' expectations for VR authentication in relation to \textbf{RQ1} and \textbf{RQ2}?

\end{itemize}

To answer these questions, we leverage \textit{technology probes}~\cite{hutchinson-chi03}, where proof-of-concept interfaces uncover hidden phenomena in user interaction, to study user authentication in VR. We designed four probes pertaining to authentication interactions for payment authentication VR -- two variants of entering a PIN, tapping a virtual card, and signing a signature -- to evaluate the user interaction experience and perceived security. 
We embed these probes in a routine payment interaction for users when they play a VR archery game, which is an organic study context.
These probes follow three interaction paradigms of authentication using something you know (e.g., PIN), something you have (e.g., token), and something you are (e.g., biometrics)~\cite{ross-nist19}. The probes and the payment context of our VR game allow us to draw valuable insights into user perception of authentication in VR.

We conducted a user study with 24 participants and evaluated their experiences in the VR game with the probes. We qualitatively analyzed open-ended responses from the participants, which is also supported by quantitative ratings, e.g., the overall usability of our designs. Our analysis reveals these findings in response to the three research questions:
\begin{itemize} [noitemsep]
\item \textbf{RQ1:} The interaction experiences of participants were associated with the perceived usability of authentication and their experience in the gamified context of VR payment. Participants benefited from intuitive virtualization and seamless interactions in authentication. However, they faced unique challenges, such as motion control. Our participants exhibited feelings of high presence and engagement in the VR game and payment. At the same time, they were sensitive to the interruptions caused by payment authentication.  

\item \textbf{RQ2:} Participants found value in realistic interactions in VR authentication as they could transfer their real-world understanding to the virtual environment. However, usability challenges and limited knowledge about VR authentication jointly hindered this translation, such as losing the sense of ownership of their signature in VR. Additionally, the immersive nature of VR heightened participants' uncertainty about threats in both the virtual and physical realms. Moreover, the gamified VR context may decrease participants' sensitivity to security risks associated with payment authentication.  

\item \textbf{RQ3:} While participants prioritized usability, secure authentication remained a crucial consideration. However, they expressed contradictory expectations from a VR authentication method. These contradictions stemmed from a mismatch between perceived and actual security, the inability to detect threats in both VR and the physical worlds, and the difficulties in accurately assessing risks in playful VR experiences.

\end{itemize}

Based on our findings, we propose recommendations to meet participants' expectations for VR authentication. These include (1) exploring virtualized interaction metaphors to bridge the gap between perceived and actual security, (2) implementing system support to detect threats in both VR and the real world, and (3) enhancing security risk communication through multiplexed feedback channels. We also highlight the open challenges and research opportunities associated with our findings and study setup, such as the optimal approach to aid security decision-making for different user groups.

\section{Related Work}
\paragraph{User authentication in virtual/augmented reality (VR/AR).} Prevailing user authentication methods on smart devices, such as smartphones, use one or more of the following factors: (1) unique knowledge (e.g., PIN or unlock patterns~\cite{von2013patterns}), (2) tokens (e.g., a device with coded ID data~\cite{nguyen-sensys16}), and (3) behavioral and physical biometrics (e.g., gestures and iris~\cite{liu2017usability, kumar-pr10}). In the context of VR/AR, authentication schemes build upon these methods but offer unique security properties compared to real-world authentication. For knowledge-based authentication in VR/AR, virtual PINs displayed in the 3D space can make shoulder-surfing more difficult~\cite{mathis2021fast}. Meanwhile, multiple input modalities can be used to select PINs or draw unlock patterns, such as eye gaze, head pose, controller tapping, and foot movements~\cite{watson-chi22, olade-icvars20}. Biometric authentication, particularly behavioral biometrics, is a prominent area of research in VR/AR. It leverages the multi-modal input modalities to capture users’ biometric traits, such as motion trajectory~\cite{kupin-icmm19}, electromyography~\cite{chen2021user}, eye tracking~\cite{zhu2020blinkey}. Behavioral biometrics are often associated with particular tasks users perform in VR/AR, e.g., throwing a ball~\cite{liebers-chi21}. There are also preliminary efforts in exploring token-based authentication mainly for AR, e.g., QR codes. User authentication in VR/AR often requires active interaction with the VR/AR interfaces. This raises usability issues and presents a tradeoff between security and privacy~\cite{stephenson-sp22}.

\paragraph{Security and privacy perception.} Prior research investigated how users perceive the security and privacy of user authentication mechanisms, with a focus on established methods like FIDO2 authentication. Lyastani et al.~\cite{lyastani2020fido2} discovered that users express concerns about security issues related to the loss of authentication tokens. Lassak et al.’s study~\cite{lassak2021s} identified misconceptions among users regarding the storage of biometric data in FIDO2 biometric authentication. These studies highlight the disconnect between users' security perception and the actual security provided by authentication methods.  

Recent research has also examined users’ security and privacy perception in VR. VR developers and users felt the lack of privacy due to opaque data collection policies~\cite{adams-soups18}. Many users center their concern around the threats from other users, e.g., as a bystander~\cite{de-csur19}. Additionally, users are worried about potential deception by digital content in VR~\cite{lebeck-sp18}. Users’ security and privacy perception of VR authentication received more attention recently. Stephenson et al. discovered that users often thought VR/AR authentication was as secure as other platforms in their online survey~\cite{stephenson-sp22}.  

Users’ perceptions of security and privacy are associated with multiple factors, including their interaction with the system. For instance, Distler et al.~\cite{distler2019security} studied how the user interface (UI) designs impact users’ perceived security of mobile e-voting apps. They discovered that inadequate UI feedback and contextual information reduce users’ sense of security. Users’ security and privacy perceptions also depend on other factors, such as personal experience. Jeong and Chiasson~\cite{jeong-chi20} found that children and adults have different interpretations and perceptions of security warnings, e.g., the symbolism of a police officer icon. Differing preconceptions are challenging for establishing trust with the system, even with visual security clues~\cite{stransky-soups21}.

\paragraph{User authentication for payment.} 
User authentication plays a crucial role in payment services by preventing fraud and minimizing financial risks. The authentication requirements vary depending on the context, such as using chip cards for in-store transactions or requiring one-time passwords (OTPs) for online shopping~\cite{acharya2013two}. Users' perception of authentication security significantly influences their adoption and use of payment services. Mobile payments have gained popularity due to users associating perceived control and security with user authentication on their device~\cite{zhang-mdpi19}. Similarly, some users also desire enhanced security using biometrics in cryptocurrency wallets~\cite{voskobojnikov-chi21}. Trust is impacted by users' understanding of how payment services ensure authentication security, such as password confidentiality~\cite{zhou-id11}. Different authentication processes in various geographies can lead to differing security perceptions~\cite{george-ndss17, busse-eurospw20}. The payment environment also affects security perception, with users considering ATM authentication riskier than payments in a restaurant due to their unawareness of attacks~\cite{volkamer-soups18}. In addition to security, factors such as usability in using a user authentication method also impact the use of associated payment service~\cite{kujala2017role}. 

\paragraph{Contributions to the literature.} To the best of our knowledge, our work is the first to systematically study the interplay between users' interaction experiences and security perception of VR authentication. Prior studies indicated such interplay in other contexts. For example, Khan et al. studied how interruptions of implicit authentication affect users' sense of security~\cite{khan-soups15}. As an emerging interaction technology, VR brings novel interaction experiences and security issues to its users~\cite{adams-soups18, de-csur19}. Thus, it is natural for us to hypothesize that interactions with user authentication in VR play a distinct role in shaping users' security perceptions. However, prior work on user authentication in VR has primarily focused on different research questions, such as enhancing its security properties~\cite{mathis2021fast, kupin-icmm19,chen2021user,zhu2020blinkey,liebers-chi21} or studying the usability and security perception~\cite{george-ndss17, stephenson-sp22, abdelrahman-avi22} of authentication respectively. Prior work in VR showed preliminary findings, e.g., users' security behaviors changed regarding their virtual surroundings~\cite{mathis-vr22}. These findings motivate our in-depth investigation of this interplay. Unlike prior studies based solely on surveys or interviews~\cite{stephenson-sp22}, our study method helped us attain in-context insights by employing ``technology probes''~\cite{hutchinson-chi03} for authentication in a realistic VR payment use case. We will also discuss how our findings supplement prior studies.

\section{Study Method}
We use a ``technology probe'' approach to understanding how interaction experiences affect participants’ perception of authentication in VR and extract design guidelines~\cite{hutchinson-chi03}. The idea of the ``technology probe'' approach entails using a set of proof-of-concept interfaces. As these interfaces package basic interactions, researchers can reveal phenomena otherwise hidden from user interactions~\cite{chandrasekaran-soups21}
This approach is commonly used to evaluate emerging technologies, including VR and user authentication~\cite{mathis-vr22,tang-chi21}.
Our probes implement the core interaction patterns of user authentication to elicit user responses in regard to their interaction experience and perception of security. We deploy our probes for participants to make routine payments in a VR game, which provides a context that mimics VR payments in real-world scenarios.

\subsection{Authentication Probes}\label{sec:probe design}
Here, we describe the process that led to the design of four authentication probes in VR.

\subsubsection{Design process.} 
Our experiment design primarily aims to create a realistic payment authentication scenario to capture authentic reactions from the participants. Toward that end, we iteratively developed our design by considering the usage context and probes together. Our research team conducted regular meetings and tests to discuss, test, and refine our concepts and implementations. 

To explore the probe design space, we drew insights from research literature and industry products, many of which are still in the early stage of commercialization. We identified three design dimensions of VR authentication: providing something you know, showing something you have, and proving who you are~\cite{ross-nist19}. These dimensions cover unique interaction patterns and inherent security properties. By including these variations, our probes helped us maximally capture different user interactions and identify common themes. We then narrowed the design scope down to authentication based on PIN, token, and behavior (gesture). 

Rather than replicating authentication interfaces in other digital contexts such as mobile apps, we designed the probes to establish a mapping of real-world experiences into VR. Such mapping has already become a popular design choice in other applications of VR for authentication~\cite{argelaguet-cg13}. To provoke participants, we opted not to optimize the usability of probes for everyone in our implementation. We, however, made our design \textit{flexible} by allowing varying gestures to interact with the interface and modification of inputs.

Our context for authentication initially involved payment at a vending machine that dispenses virtual objects. However, we found, in pilot studies, that this task was not ideal for engaging people in VR. As a result, we shifted our focus to integrating payment authentication within VR games, a more popular scenario. Building on this context, we further refined our probe design of payment authentication to align with the virtualization of this context.

\subsubsection{The Four Probes.}

\begin{figure*}[h]
\centering
\includegraphics[width=\textwidth]{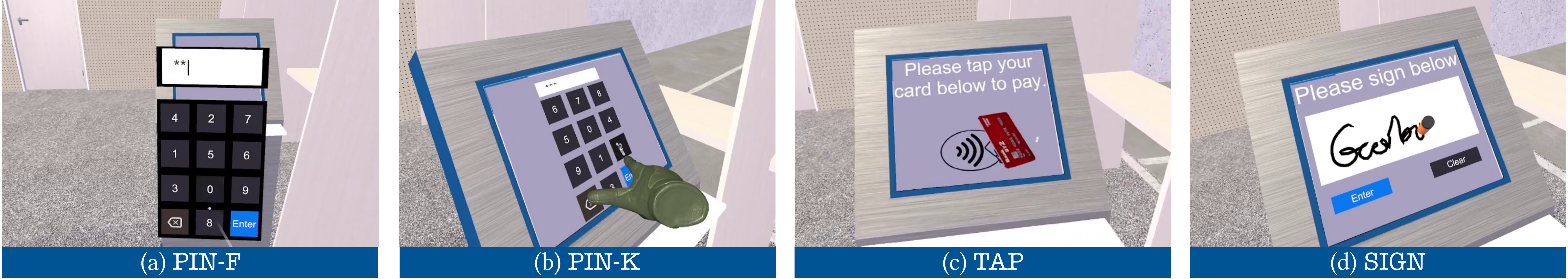}
\caption{Our four probes for VR authentication.}\label{fig:Probe_design}
\end{figure*}

We designed four probes: (1) floating PIN pad (\pinf), (2) on-kiosk PIN pad (\pink), (3) tap-to-pay (\tap), and (4) signature (\sign). We now describe these four probes, which are shown in Figure~\ref{fig:Probe_design}.

\paragraph{$\bullet$ Floating PIN pad (\pinf).}
\pinf resembles the default virtual input interface of many VR platforms, where participants interact with a floating PIN pad. \pinf also conceptualizes the idea of giving participants a personal and isolated virtual experience in the authentication.  Participants enter PINs by pointing to a floating PIN pad that follows the participants’ viewport.  \pinf randomizes the PIN layout, which is a common security mechanism in digital PIN pads against observers~\cite{mathis-vr22}. Note that \pinf serves more like a baseline authentication probe.

\paragraph{$\bullet$ On-kiosk PIN pad (\pink).}
In contrast to \pinf, \pink presents a better mapping of physical-world experiences by rendering its PIN pad on the kiosk where participants initiate and confirm payment. As such, participants do not experience a gap in the transition between authentication and other payment tasks, e.g., selecting the items.  

\paragraph{$\bullet$ Tap-to-pay (\tap).} 
\tap represents how participants could own and use a personal virtual object as a unique token to authenticate. To pay, participants take out a virtual credit card from an inventory on their virtual body and tap the card on the kiosk display. The kiosk checks whether the card is in proximity and being held by the user. 

\paragraph{$\bullet$ Signature (\sign).} 
\sign represents how participants draw a unique signature to give consent and prove their identity. It virtualizes the real-world signing processes.
Participants grab a virtual pen from the kiosk and sign in the designated area using the hand-held controller; Only after participants sign, they can proceed to check out.\\

Our design achieves the goals of the technology probe method to collect in-context information about the usage, test out the technology, and inspire future design by upholding the principles as proposed by Hutchinson et al.~\cite{hutchinson-chi03}. As the functionality of technology probes should be simple, our designs only package the essential frontend interaction instead of the full backend of authentication mechanisms, e.g., verifying an encrypted token or comparing signatures. Instead, our study adopts the idea in ``Wizard-of-Oz'' studies~\cite{mecke-icmum18} in creating an impression that the system has full functionality through a mock registration process. Also, our probes also support logging of participants' interaction with the interface, including the fine-grained timing, to supplement the analysis.

\subsection{Experiment Context: VR Archery Game}
\label{sec:game_design}

\begin{figure*}[t]
\centering
\includegraphics[width=\textwidth]{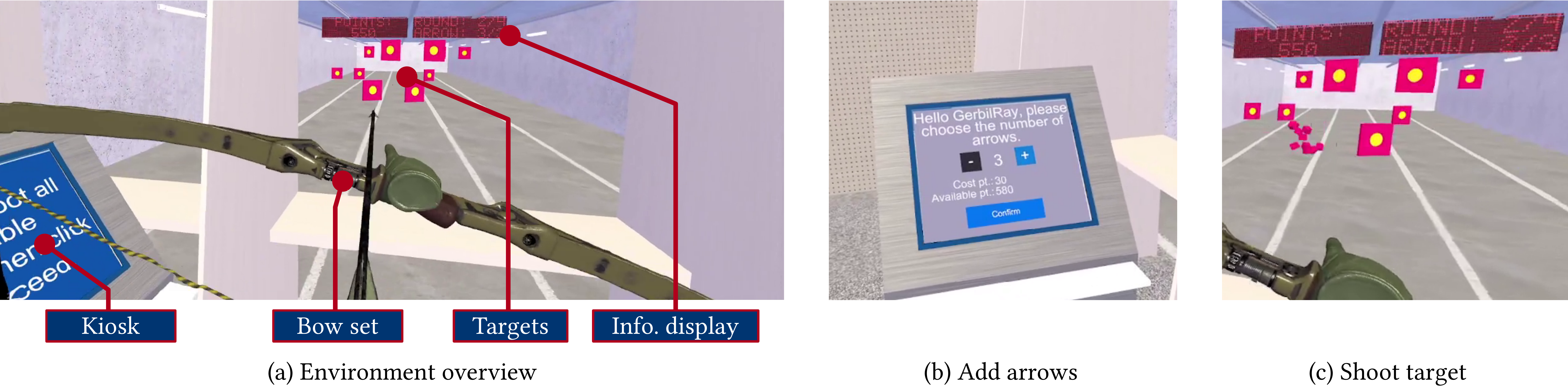}
\caption{The environment of our archery game with key components marked and key scenes illustrated. }\label{fig:game_scene}
\end{figure*}

We design a VR game -- an archery contest -- where participants trade in-game credits, using the probes to authenticate their payment. Participants earn credits by shooting virtual targets and consume these points when they refill arrows. 
To make the game and payment realistic to participants, we match their in-game credits with a physical compensation: the participant who scores the highest wins a ``grand prize'' (a small gift).
Such compensation, which is added to our base compensation for participating in the study, is standard in prior studies that include gamification~\cite{brewer-idc13, mccoy-18, tondello-19}. We determined the value of the prize (a 90 USD fitness tracker) based on our local wages. Below, we describe the main components of this game (see Figure~\ref{fig:game_scene}).

\paragraph{Environment.} We situate the contest in an indoor archery range. We assign the participant a chamber. In the chamber, the participant can find the bow set and a kiosk instructing them to shoot, refill arrows, and authenticate their payment. We place the targets towards the other end of the room at a distance. The participant can also find the information display that shows their current credits and remaining progress. 

\paragraph{Bow set.} The participant can interact with the bow using the two hand-held VR controllers. They need one hand to hold the bow and the other hand to draw. The participant shoots the arrow toward the target by releasing the drawing hand. The participant will need to refill after they use up three arrows. Each arrow consumes ten credits.

\paragraph{Targets.} We place eight targets in the range at two different distances. Taking down each closer target rewards 20 credits to the participants while 30 for the farther targets. 

\paragraph{Kiosk.} The kiosk displays game instructions and payment interfaces. To refill arrows, participants interact with the kiosk to select how many arrows to load. After the participant confirms the selection, the kiosk will display the authentication interface among one of the four probes. Once the participant completes authentication, the kiosk will display a waiting page that emulates the running backend processes. After that, the kiosk will accept the credit payment and refill arrows. 

\paragraph{Information display.} The display shows the current credit, the remaining arrows, and how many rounds the participant has completed.

\paragraph{Implementation.}
We implemented our VR game using Unity, a mainstream VR engine, and C\#.
In the experiment, we ran the game on a commodity PC (CPU: Intel i7-12700F, GPU: Nvidia 3060Ti), which is connected to an Oculus Quest 2 VR headset.
Participants interacted using the headset and its hand-held controllers.

\subsection{Experiment Design}
Here we explain our experimental procedure, the instruments we used to collect participants' responses, and our recruitment.
\subsubsection{Experimental Procedure.}\label{sec:expreiment_procedure}

\begin{figure*}[t]
\centering
\includegraphics[width=\textwidth]{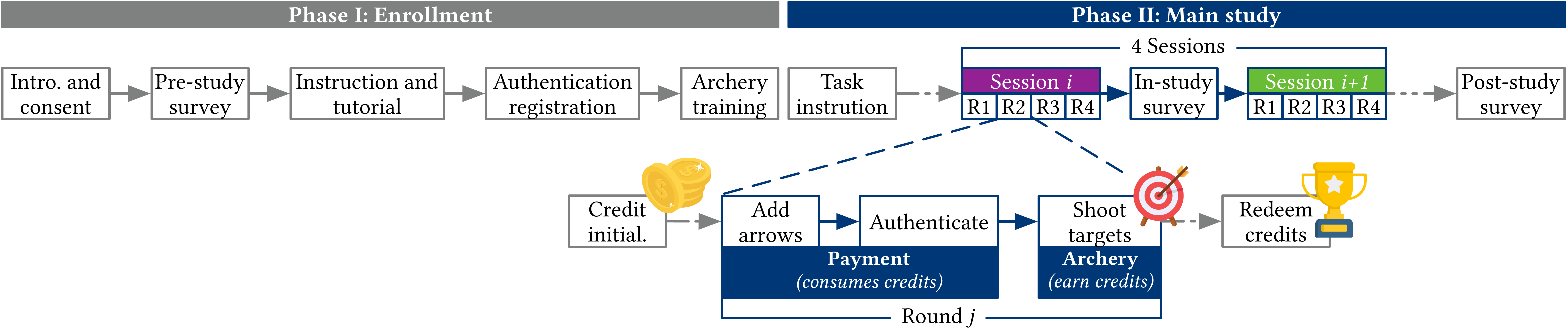}
\caption{Our experimental procedure.}\label{fig:game_flow}
\end{figure*}

We designed a with-in-subject experiment to evaluate the four probes.
The study consists of two phases--an enrollment phase and the main study phase.
The purpose of enrollment is to familiarize participants with our VR and authentication setups. In the main study, participants played the VR game and interacted with our probes during authentication.
Figure~\ref{fig:game_flow} illustrates our experimental procedure.

\paragraph{Enrollment.} The experimenter first obtained consent and asked the participant to complete a pre-study survey. After that, the experimenter explained the game context to the participants. We assigned participants a ``nickname’’ for the game. The experimenter then introduced the PIN, the virtual card, and the signature setup to the participant (we assigned each participant the same credential information for the statistics of authentication time). We described to the participants that they would enroll the information, including the signature, for the use of authentication in the second phase. 
The experimenter then walked participants through how to use our VR setup.
After the tutorial, the experimenter let the participants practice with the registration kiosk, similar to the one they will use for authentication, and present their credentials to it.
After registration, the participant will practice archery. The enrollment phase took about 20 minutes.

\paragraph{Main study.} The one-hour main study took place on another day to reduce participants’ fatigue.
We first reminded the participants of the study procedure, game context, and the authentication interaction.

Then, the participants entered the game with 400 points. They proceeded to finish four sessions; each included the game with an in-study survey. In each session, the participants paid using one of the probes (in a randomized order) to authenticate. Each session consisted of three rounds. Participants authenticated to pay for the arrows (min: 1, max: 3), except for the first round of every session where three arrows were given for free. At the end of each session, the participants completed an in-study survey, and following the second session, they were given a brief break outside of VR. Once all four sessions were completed, participants were instructed to fill out a post-study survey. After the completion of the survey, participants redeemed their credits to compete for the grand prize.

After the experiment, we disclosed our full study purpose to the participants. We clarified that our focus is to evaluate the interaction experiences and perceptions from the frontend, and we did not process their information, i.e., signature, in the backend.

\subsubsection{Instruments.}
\label{sec:Instruments}
Our study mainly relies on surveys that elicit participants' responses for our qualitative analysis. 
We chose to use surveys instead of more active methods such as think-aloud studies~\cite{charters-bej03} to minimize the interference with participants' game experiences.

\paragraph{Pre-study survey.} In the pre-study survey, we collected participants' demographic backgrounds, including their gender, age, education, and experience with VR. We also used the standard affinity for technology interaction (ATI) score (computed from 6-point Likert ratings) to understand participants' tech-savviness~\cite{franke-ijhci19}.

\paragraph{In-study survey.} The in-study survey assessed (1) participants' general experience in the VR context and (2) the perceived usability of authentication probes.   
For the former, we used the IPQ VR presence questionnaire, a standard measure of participants' sense of presence in VR. 
The sense of presence is a commonly adopted measure of VR experience. It is defined as participants' subjective perception of being and acting in the virtual world (though their body resides in the physical world)~\cite{schwind-chi19}.
This measure consists of four sub-scales (1) sense of being here (PRES), (2) spatial presence (SP), (3) involvement (INV), and (4) experienced realism (REAL).
For the perceived usability, we collected participants' open-ended comments on the usability issues of each authentication probe when participants exited VR. Moreover, participants completed the system usability scale (SUS), a standard measure to assess the overall perceived usability, for each authentication probe~\cite{brooke-jus13}.

\paragraph{Game log.} In addition, we logged participants' behaviors and timings in the game, including their archery performance as well as the time spent on authentication.
We used these objective behavior logs to support our qualitative findings.

\paragraph{Post-study survey.}
The post-study survey consists of three major components. 
First, to understand participants' engagement in the routine payment of the game, we asked them to explain how they decided on the number of arrows to pay. 
Second, we wanted to understand participants' security perception of payment authentication.
We designed questions to elicit participants' responses from different angles. 
From prior research, we identified the five aspects related to the security of payment authentication, namely \textit{consent}~\cite{herzberg-cacm03,lyastani-sp20}, \textit{security}~\cite{kim-ecra10},  \textit{privacy}~\cite{zimmermann-ijhci20}, \textit{being alerted}~\cite{reese-soups19,khan-soups15,wolf-chi19}, and \textit{in control}~\cite{nilsson-chi05}.
We asked the participants to evaluate and elaborate on their agreement on statements related to the five aspects. One example is: when the participant used \tap to pay for arrows, ``I [the participant] felt that my [the participant's] payment was secured''.
We collected both their Likert-scale rating and open-ended responses to explain perceptions. 
However, rather than relying on the quantitative ratings, we mainly study the relation between interaction experiences and perception from their open-ended responses.
Third, the post-study survey asked participants about their preferences among the four probes, their quality expectations that affect their preferences, and their suggestions to improve these probes.
We use these questions to further understand participants' expectations of VR authentication.

\subsubsection{Participants and Recruitment.}
Consistent with prior work in VR authentication~\cite{mathis-vr22}, we recruited 24 participants from our organization. We stopped recruiting when we observed data saturation from our qualitative coding~\cite{guest-fm06}. Each participant received compensation (a 30 USD gift) after they completed the study.

The demographics of participants are as follows. 18 out of 24 identified themselves as man (6 women). The average age of our participants is 29.6 years old (std: 4.8). 20 of them completed or were studying for a graduate degree. Most participants have a background in computer science. Participants' ATI score (mean: 4.35, std: 1.1) shows a high affinity for technology interaction)~\cite{franke-ijhci19}. 20 participants had used VR before (mainly for gaming). But none of the 20 participants frequently used it.

\subsubsection{Ethics and participant safety.} Our study and recruitment were approved by the IRB-equivalent body of our organization. While the study poses a low risk to participants, we took the following steps to protect their physical safety and data privacy. First, we communicated potential risks and their right to withdraw through our informed consent and disclosure processes. Second, we spread our study into two days to minimize fatigue. In addition, we regularly checked in with the participants and made sure they were comfortable to proceed during the study. Third, we ensured that our physical space was clear and safe for the use of VR. The VR application notified our participants when they got close to physical boundaries. Last, our data collection and analysis do not include sensitive personal information, and all data were anonymized and stored in a secure server in our organization.

\subsection{Data Analysis}
We analyzed survey and log data as described in Section~\ref{sec:Instruments}. To analyze such qualitative data, the first author started open coding and took memos while recruiting participants. Meanwhile, another researcher coded data independently. We coded each response the participant made corresponding to a question. The whole team discussed the memos, reconciled the codes, and refined the codebook iteratively. The two coders reached high inter-rater reliability (Cohen’s Kappa $\kappa$ = 0.81) using responses from 6 randomly sampled participants for each question (78 out of 312 responses in total), then we converged on a codebook to code the rest of the data.
Using Grounded Theory~\cite{walker2006grounded}, high-level themes emerge from our coding.
When we observed data saturation from our coding~\cite{saunders2018saturation}, we stopped recruiting. 
We make our codebook available via an anonymous link.\footnote{Our codebook and survey are available at: \url{https://osf.io/re9py/?view_only=e2e01010cf484bd8b476442a0c92398b}}

\begin{figure*}[ht]
\centering
\includegraphics[width=\textwidth]{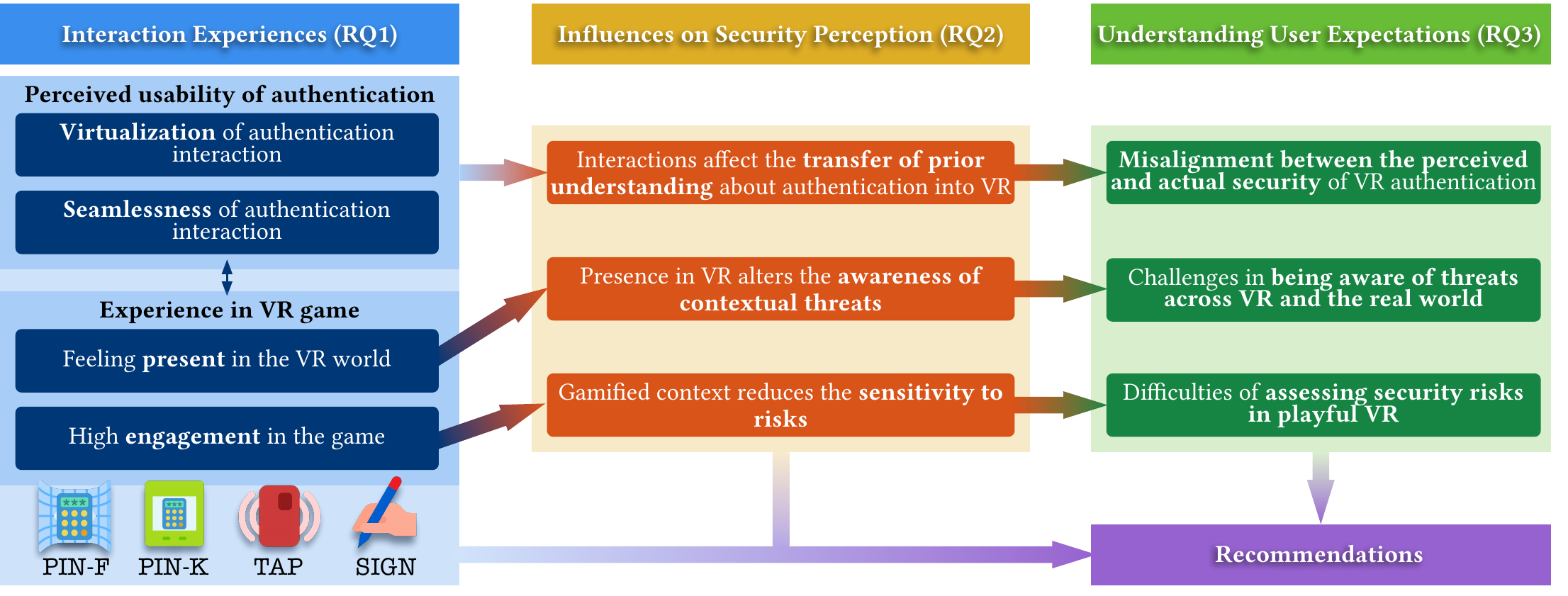}
\caption{Our analysis framework and summary of results. We connect themes that are most related across research questions.}\label{fig:framework}
\end{figure*}

In Figure~\ref{fig:framework}, we show our analysis framework along with themes we identified from the analysis.
In addition, we report the quantitative data to support our qualitative analysis, including the IPQ questionnaire, SUS scales, and participants’ Likert-scale rating on security and privacy perception.

\subsection{Limitations}

Our study has the following limitations. First, our participant population has a demographic bias, e.g., most participants are tech-savvy. Our population and demographics are comparable with prior work applying similar methods~\cite{ohagan-chi23, yan-imwut20,liang-imwut20, mathis-vr22}. Using a technology probe, our study's objective is not to generalize but present a set of findings and recommendations that guide future design.
Nevertheless, our analysis reveals that even these tech-savvy participants faced challenges in interacting with VR and assessing security, let alone other users. 
Future work may study the probes with a more diverse population, generalize the findings, and quantitatively measure users' experiences and perceptions. 
Second, our study does not investigate how users use VR authentication longitudinally in the wild where users’ experience and perception of security may change over time. In addition, participants' self-reported responses may be biased due to social desirability~\cite{phillips-ajs72}. Last, the purpose of our study is to understand users with probes in an early design phase. We focus on the novel experience of fundamental authentication concepts in VR payment. As such, we did not explore all alternative implementations of the VR interaction and interface.
Despite these limitations, we believe that our work still presents a significant contribution. To the best of our knowledge, our exploratory study is the first to investigate the interplay between interaction experience and perception of security for VR authentication.

\section{RQ1: Interaction Experiences}

We observed two overarching themes of participants' experiences that influence their security perception of authentication. The first is the \textbf{perceived usability of authentication}, which consists of two sub-themes: \textit{virtualization} and \textit{seamlessness} of the authentication interaction. The second theme is \textbf{experience in VR game context}, which is influenced by the two sub-themes: participant \textit{presence} and \textit{engagement}. In the following, we elaborate on these themes and discuss how our probes uncovered these experiences.

\subsection{Perceived Usability of Authentication}
The virtualization and seamlessness of the authentication interaction affect the perceived usability. Our probes entail additional aspects regarding \textit{authentication interface}, \textit{process of authentication}, and \textit{motion control} that are not typically present in real-world authentication interaction.  These aspects contribute to the virtualization and seamlessness of the VR authentication, leading to different perceptions of usability. 

\subsubsection{Virtualization of Authentication Interaction.}
Participants expressed positive feedback regarding the usability of \textit{intuitive} VR interactions and interfaces that resemble their real-life experiences. Virtualizing \textit{familiar} interfaces from the real world made them feel more \textit{immersed}. However, complex VR interactions introduced an additional learning curve, particularly with \sign, which involved more friction compared to PIN methods and \tap. For instance, one participant felt \textit{``signing in the virtual world was very different as compared to the real world''} (P13). Nevertheless, another participant found \sign was easy after practicing. %
\begin{displayquote}
\textit{``It's very easy to learn and use, and the functionality can be easily picked up.''} (P11)
\end{displayquote}

An intuitive and realistic \textit{authentication interface}, in terms of presentation and feedback, can enhance the virtualization of interaction, even though the authentication concept is new to participants. The token-based authentication \tap was favorable to participants, primarily due to its interface presentation:

\begin{displayquote}
\textit{``since it closely mimics the way I use card payments in my day-to-day life.''
} (P8)
\end{displayquote}

Participants exhibited different preferences over the interaction modality, often by comparing it with real-world counterparts. One participant mentioned, when describing the virtual PIN pad, that the \textit{``laser pointer is less similar to using keyboards in real world''} (P2). 
Moreover, participants expected more realistic interaction feedback like what they receive in the real world, especially for the highly interactive \sign approach. For example, they demanded a better sense of writing on a ``paper'' in VR.
\begin{displayquote}
\textit{``It is still usable and the pen writes like real writing but still did not get the feeling of writing on a paper.''} (P1)
\end{displayquote}
In addition, participants perceived virtualization differently due to their prior experience in the real world, i.e., how they performed payment. For example, one participant was not used to a shuffled PIN pad as \textit{``most terminals have a fixed layout (I am [they are] thinking of ATMs and gas stations)''} (P5).

Generally, the virtualization of real-world authentication interactions helped improve perceived usability. However, some participants expected VR interfaces to save them more effort compared to authentication interactions in the physical world, e.g., moving \textit{``to get a better look at the numpad''} on the kiosk (P13).

\subsubsection{Seamlessness of Authentication Interaction.}
The seamlessness of authentication interaction also contributed to better-perceived usability. Interestingly, participants associated seamlessness with specific aspects such as the \textit{comfort} of interacting in VR, the \textit{physical, mental, and time efforts}, and the \textit{smoothness} of the interaction process.

As mentioned earlier, our VR authentication probes exhibit familiar properties of real-world authentication. They also inherit some of the difficulties from previous authentication experiences, including the transitions in the workflow, the memorability of PIN, and the additional usability challenge to use a shuffled PIN pad where \textit{``the numbers shifted locations between attempts''} (P13). 

The \textit{motion control} in VR introduced unique challenges that made authentication interaction less seamless. This aspect had a more noticeable impact on probes that required complex interactions (\sign, \pinf, \pink) compared to \tap.  The challenges with motion control included inconsistency in action control when interacting with digital components. This inconsistency affected the accuracy of translating participants' actions, such as how their virtual gestures aligned with their intentions. For example, one participant preferred \pink over \pinf due to the observation that \textit{``stable kiosk was easier to use as it allowed for better calibration''} in entering PINs (P7). 

Moreover, spatial awareness in VR, such as coordinating the movements of a virtual object and the avatar, contributed to the challenge of motion control. Some participants indicated that lacking spatial awareness hindered their ability to control a virtual object, e.g., stretching their arm to sign a signature:
\begin{displayquote}
\textit{``It takes a while to get used to the proper distance between the pen and the kiosk screen.''} (P7)
\end{displayquote}

These aspects became evident when participants commented on the seamlessness of the \sign probe. As mentioned before, some participants thought \sign was easy, while others did not. One participant felt that performing the VR gesture was easy conceptually but hard in practice.
\begin{displayquote}
\textit{``Signature: it was conceptually easy, but the execution was somewhat cumbersome and it required a complex gesture.''} (P6)
\end{displayquote}

\subsubsection{Quantitative Usability Analysis.} 
\begin{figure*}[h]
    \begin{minipage}[t]{0.4\linewidth}
        \centering
        \includegraphics[width=\linewidth]{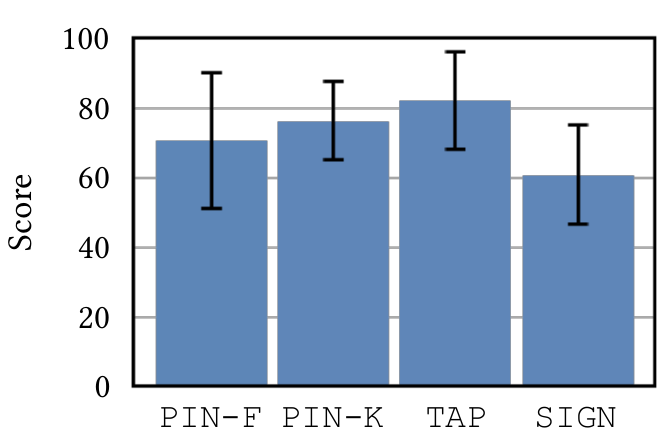}
        \caption{SUS scores (average and standard deviation) for each authentication probe.}
        \label{fig:sus}
    \end{minipage}%
    \hfill
    \begin{minipage}[t]{0.56\linewidth}
        \centering
        \includegraphics[width=\linewidth]{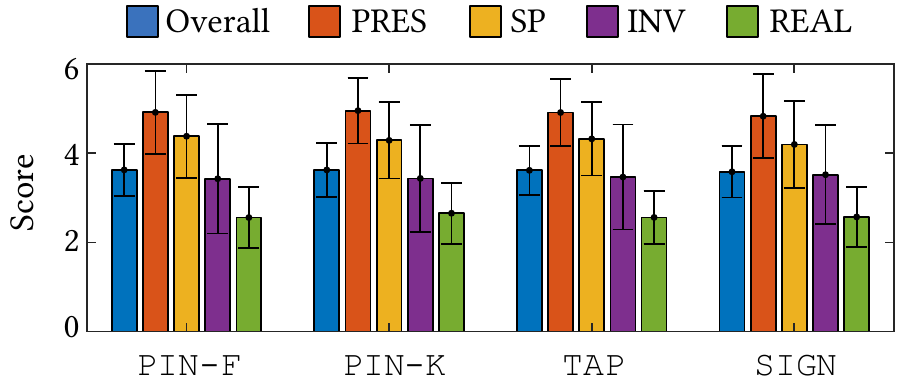}
\caption{IPQ sense of presence scores for game contexts with the four authentication probes. A higher score indicates a higher level of presence. From 0 to 6, a score higher than 3 stands for neutral. All subscales have a positive mean score, except REAL. The IPQ scores of our setups are consistent with prior implementations for VR authentication~\cite{mathis-vr22}.}\label{fig:ipq}
    \end{minipage}
\end{figure*}

The above interaction experiences manifest in different usability ratings for the four probes. By comparing participants’ SUS rating (Figure~\ref{fig:sus}), a standard usability metric, we find that \tap received the highest SUS score (mean: 82.1, std: 14.6) and cost the least time (mean: 3.3s, std: 2.0s), indicating ``excellent'' usability~\cite{brooke-jus13}. It benefits from seamless and intuitive interaction.
\pinf (mean: 70.6, std: 19.9) and \pink (mean: 76.1, std: 11.7) closely follow \tap, demonstrating a ``good'' usability. They consumed comparable time to complete: \pinf (mean: 10.8s, std: 11.1s) and \pink(mean: 10.4s, std: 6.6s). As discussed, participants appreciated the usability of these methods due to their familiarity and relative ease, despite the effort required to memorize and enter the PINs.  Moreover, the differences in scores between \pinf and \pink are small. Participants commented on different usability issues for them, such as the inconvenience of walking towards the kiosk for \pink and the distraction caused by entering PINs on a moving panel for \pinf.
\sign (mean: 60.7, std: 14.7), which is also the most time-consuming probe (mean: 18.3s, std: 9.8s), received the lowest score among the four probes but still achieved ``OK'' usability. Although its interface appeared intuitive, signing with the virtual pen proved to be challenging for multiple participants in the VR environment.

\subsection{Experience in the VR Game Context}

In addition to the interaction experience with the authentication probe, participants' experience with the VR game (as a proxy for the VR context) appeared to influence their perception of security. We observed two sub-themes related to how participants felt a sense of \textit{presence} and \textit{engagement} in the game.

\subsubsection{Feeling Present in the VR World.} 
Using the IPQ presence questionnaire (Figure~\ref{fig:ipq}), we observed that participants rated their presence in the VR world positively. However, we did not observe a significant difference in the IPQ scores between sessions with different authentication probes, which is consistent with previous findings about authentication is often perceived as a secondary task~\cite{de-soups10}. When participants felt present in this world, some of them expressed dissatisfaction when authentication disrupted their immersive experience through an ``unreal'' interface, e.g., \pinf:
\begin{displayquote}
\textit{``the floated pad makes it so unreal that I know it is in VR rather than real life. I do not like the experience.''} (P24)
\end{displayquote}

\subsubsection{High Engagement in the Game.} 
We observed the participants to be highly engaged in the archery game and the routine payment. When deciding which arrows to pay for, participants mentioned various factors, including  \textit{their strategy to compete} and \textit{enjoyment of the game}. Among the participants, 19 explicitly mentioned that they would like to shoot as many arrows as they could, and six people said it was fun to play, despite differences in their archery performances (highest score: 1470, lowest score: 190). This high level of engagement also led some participants to expect less effort during authentication compared to the time spent actively playing the game, as expressed by one participant: \textit{``the time I am [they are] actually playing the game''} (P18).\\

\noindent
{\colorbox{beaublue}{\textbf{Takeaway-RQ1.}}}
Participants' interaction experiences included their perceived usability of authentication and their experience in the VR game context. Our findings were consistent with prior observations that PIN authentication has acceptable usability ~\cite{george-ndss17} and that user authentication is perceived as a secondary task compared to the overall context~\cite{mathis-vr22}. We further observed that participants associated the usability of authentication with better virtualization and more seamless interactions. When fully engaged in the VR game, participants preferred less disruptive experiences to keep their immersion. Moreover, we found that unique usability challenges of VR interactions, such as motion control, influenced participants' authentication experience, particularly with behavior-based authentication.

\section{RQ2: Influences on Security Perception}
\label{sec:RQ2}
Previously, we listed the interaction themes that influenced user perception of the security of authentication. In the following, we discuss \textit{how} these interaction themes influenced participants’ security risk assessment while authentication. Our qualitative analysis of participant responses revealed three themes of influences in Figure~\ref{fig:framework}. These themes include forming security perceptions, being aware of threats, and assigning risk to threats.

\subsection{Interactions Affect the Transfer of Prior Understanding About Authentication into VR}
\label{sec:RQ2-1}
Facing a limited understanding of novel VR authentication concepts, participants transferred their prior experience with real-world authentication to VR. This transfer of knowledge, however, did not result in a consistent security perception. The virtualization and seamlessness of authentication interactions, while enhancing usability, affected how participants formed their security perceptions of VR.

\subsubsection{Impact of Prior Knowledge and Experience.}\label{sec:RQ2-1-1}
Participants \textit{associated the virtual probes to authentication processes from the physical world with which they are familiar}. Examples of such processes include possessing secret knowledge (PIN) that is \textit{``only known by me [the participant]''} and using a shuffled PIN pad that \textit{``gave me [the participant] some sense of security''}  (P20, P17). 

However, participants' \textit{prior understanding and real-life experiences varied}, leading to differences in their perceptions. For example, the participants held different views on how consent works in VR payments, as shown in Figure~\ref{fig:overall_perception}. For example, one user viewed \sign as less familiar as they \textit{``rarely signed to pay (only recently in the US…)''} (P6). Meanwhile, others \textit{``naturally perceive it as giving my [their] consent''} as they would in the physical world (P12). Some participants associated consent with privacy implications. For instance, P20 raised a possible privacy concern with \sign, which \textit{``requires more information [signature]''} (P20) than PINs. 

Meanwhile, participants \textit{expressed varying levels of confidence about the security of authentication}. Some participants were not confident due to lacking information, for example, \textit{``without more information on how the payment actually works''} or without knowing \textit{``how the backend works''} (P13, P3). P3 also suspected that \tap would need additional verification or certification steps for the virtual card. On the contrary, some participants more confidently assumed that \textit{``the technology behind tapping makes me [them] believe it is safe''} (P14). This finding contributes to the polarized responses, e.g., security and privacy of \tap (Figure~\ref{fig:overall_perception}(a, b)).

\subsubsection{Impact of Usability.}
\label{sec:RQ2-1-2}
We observed that, because of a limited understanding of VR authentication, the usability properties of the probe affected the participants' security perception. This was the case for the \sign and \tap probes. In particular, as VR interactions do not mimic real-world sensations very well, participants felt a loss of control over authentication. For example, some participants were not confident they were providing consent when using \sign and \tap, compared to \pinf and \pink (Figure~\ref{fig:overall_perception}(a)). When signing in VR, some participants did not feel that the VR signature belonged to them. One participant expressed: 

\begin{displayquote} 

\textit{``The sign-to-pay method was a bit hard to use, so I think I just tried to write something, and I felt less like providing my signature.''} (P10) 

\end{displayquote} 

Similarly, the same participant felt that \tap was not secure as they still thought \textit{``it's not my [their] real card but a card-like object''}. In addition, the lack of feedback associated with the seamlessness of the VR interaction contributed to the loss of control, particularly for \tap. For instance, P2 expressed concern that \tap appeared too \textit{`` no-brainer ''} without any warnings (P2).

The \textit{usability characteristics of the interaction appear to undermine participants' confidence in security}. Some participants expressed concerns about the security of \sign, the authentication method with the lowest usability. They believed that this method compromised security by accepting inconsistent signatures, which could potentially make them more susceptible to impersonation.

\begin{displayquote} 

\textit{``The sign-to-pay felt the most insecure as people easily have access to my cheques and can probably fake in the VR world since the VR signatures were clearly less accurate than the real-world.''} (P5) 

\end{displayquote} 

The participants expressed greater confidence in the security of traditional PIN methods, which also offered more acceptable usability. In such cases, a higher level of interaction provided participants with a sense of control (Figure~\ref{fig:overall_perception}(e)). In a similar vein, a subgroup of participants expressed a willingness to accept additional interaction to prioritize security and transparency. This included incorporating extra warnings and utilizing a shuffled PIN pad to avoid \textit{``making accident payment''} (P15).

\subsection{Presence in VR Alters the Awareness of Contextual Threats}
\label{sec:RQ2-2}
Participants engage in threat modeling as they analyze the security properties of the VR authentication. During this process, they identified several entities based on their understanding of real-world payment and the VR environment. These entities included payment and authentication service providers, physical and virtual bystanders, third-party apps, and malware. Participants also \textit{recognized software vulnerabilities} that could lead to compromising personal information, such as PINs and signatures:
\begin{displayquote}
\textit{``Similar reason that pins and signatures are just exposed to the game or malware in the gaming system.''} (P24)
\end{displayquote}
Another example is participants' \textit{awareness of virtual and physical bystanders}. For example, P6 considered \pink less secure than \pinf if bystanders in VR were able to see the kiosk.

Participants expressed concerns that their presence in the VR environment reduced their awareness of both the virtual and physical worlds. This lack of awareness potentially exposed their information (PIN, signature) and assets (cards and tokens) to adversaries. They worried that \textit{attacks in VR could be more imperceptible than in the physical world}, such as being deceived by an invisible terminal for phishing purposes:
\begin{displayquote}
\textit{``Tap-to-pay is the easiest, but I feel it not safe because I can easily touch it to an invisible system in the VR world.''} (P19)
\end{displayquote}
\textit{Noticing attacks depended on the type of authentication interaction in VR}.  For example, it might be easy for a person to notice an attacker \textit{``stealing a card,''} but harder to observe a physical or virtual shoulder surfing attack when signing or entering a PIN (P6). Participants were also concerned that malicious users could leverage prior knowledge about the victim in the physical world to launch attacks in VR:
\begin{displayquote}
\textit{``signing can be also copied by anyone who knows my signature in the real world.''} (P12)
\end{displayquote}

Last, some participants had \textit{a general lack of trust due to their ambiguity surrounding the VR technology and authentication}. P12, for instance, expressed skepticism about the security of \tap, which stems from their distrust of the VR technology:
\begin{displayquote}
\textit{``Tap-to-pay was the simplest, but it was way too simple to believe that the entire payment process behind the scene was dealt with as I wanted. I'm not sure if this is due to my distrust to the specific payment system, or just to the VR world, or both.''} (P12)
\end{displayquote}

\subsection{Gamified Context Reduces the Sensitivity to Risks}
\label{sec:RQ2-3}
Our context for user authentication revolves around participants' routine payment in a VR game. During these activities, participants were highly engaged in the game and the payment process. Some participants noted that their \textit{sensitivity to authentication security related to the relevance of virtualization to payment}. For instance, P9 felt a greater sense of control when using \tap, as it simulated an actual payment experience by physically grabbing the card and initiating the payment. On the other hand, \pink, \pinf, and \sign were perceived as less specific to payment.

However, the \textit{gamified context may reduce participants' sensitivity to security and privacy risks}. For instance, P7 argued that they did not feel a loss of privacy in a VR game compared to real-life transactions.  
\begin{displayquote}
\textit{``In the context of the game I didn't feel like giving away privacy. However if I were to imagine this with real transactions, then I'd feel like giving some of my privacy away, similar to every time I pay with something else than cash.''} (P7)
\end{displayquote}

Furthermore, the gamified interactions and interfaces of certain authentication probes made some participants feel less attentive to security. When signing a signature in VR, the gaming aspect overshadowed P7’s sense of giving consent.
\begin{displayquote}
\textit{``The sign to pay wasn't completely obvious you were actually paying for something, it could have been part of the game to have to write your name.''} (P7)
\end{displayquote}
Similarly, P10 thought \pinf \textit{``felt a bit too much like being in a game as well''} compared to \pink (P10).

\begin{figure*}[h]
\centering
\includegraphics[width=\textwidth]{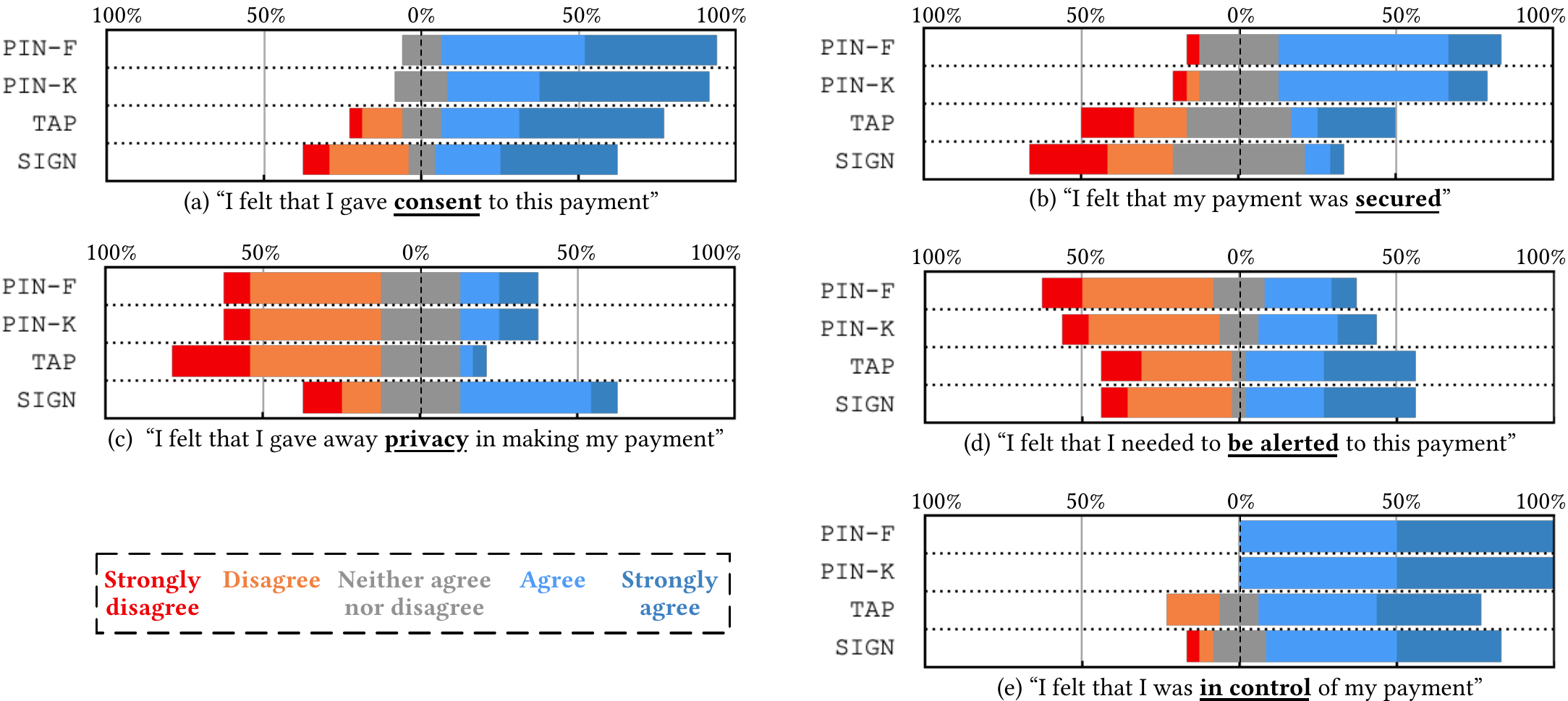}
\caption{The overall security perception of the four probes. We color-coded the bars that represent the percentages of participants (red: strongly disagree, orange: disagree, grey: neither agree nor disagree, light blue: agree, dark blue: strongly agree). }\label{fig:overall_perception}
\end{figure*}

We also observed a dichotomy in whether participants prefer to be alerted to the fact that a payment is taking place (Figure~\ref{fig:overall_perception}(d)). Some participants felt like they need to be alerted because the lack of feedback and their limited understanding make them feel less secure (as we explained in Section~\ref{sec:RQ2-1}). Other participants felt like the payment authentication was secure, and they did not see a need to be alerted if not prompted. For example, the more personal interface of \pinf captured a participant's attention because it \textit{``clearly wouldn't let me [them] proceed through the game''} (P16).\\

\noindent
\colorbox{beaublue}{\textbf{Takeaway-RQ2.}} 
Our findings confirmed our hypothesis that authentication interactions have an impact on the security perception of VR authentication, similar to previous studies on applications like implicit authentication~\cite{khan-soups15, wiefling-acsac20}. Our findings in VR authentication further complement previous work as follows.
Stephenson et al.'s survey~\cite{stephenson-sp22} mainly discovered that the usability of knowledge-based authentication in VR is perceived as a tradeoff between security and usability. Participants in our study generalized this tradeoff to other forms of VR authentication, including behavioral biometrics. Moreover, we observed that the virtualization of authentication interactions in VR may bias participants when transferring prior understanding of authentication to VR, potentially leading to a false yet confident sense of security. Our participants also expressed uncertainties regarding threats in both the virtual and physical realms due to the immersive nature of VR. Additionally, the gamified context of VR had the potential to reduce participants' sensitivity to security risks.

\section{RQ3: Understanding User Expectations}
\label{sec:RQ3}
\begin{figure*}[t]
\centering
\includegraphics[width=0.5\textwidth]{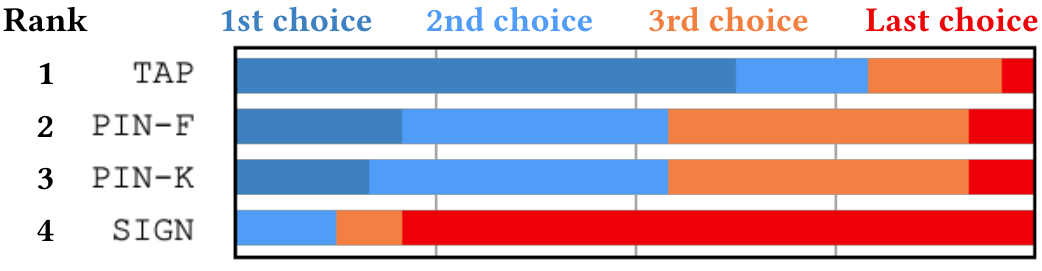}
\caption{Overall payment preference of the four probes. The bars are color-coded dark blue, light blue, orange, and red to indicate the fraction of participants that selected each probe respectively to be their 1st, 2nd, 3rd, and last choice.}\label{fig:payment_pref}
\end{figure*}

After identifying how interaction experiences shape the security perception of VR authentication users, we turn our attention to participants' expectations of an ``ideal'' authentication  experience. Our subsequent discussion builds on our findings in \textbf{RQ1} and \textbf{RQ2} as well as qualitatively analyzing participant responses to the surveys. 

\subsection{Misalignment between the Perceived and Actual Security of VR Authentication}

User authentication in VR is a new concept for users, where \textit{a gap still exists between the perceived and actual security.} The technical properties of VR authentication mechanisms may differ between the real world and VR, despite the similar virtualization metaphors. For example, the technology to secure a VR token would be different than the EMV chip in a physical card even though they share a card metaphor. We observed that some participants, even those who were technologically savvy, tended to overestimate the security of VR. On the other hand, we found that making the authentication more oblivious and seamless to users may make them underestimate the actual security in VR (Section~\ref{sec:RQ2-1-1}). The observations are associated with users' lack of understanding of these novel technologies in VR to assess the actual security. For example, one participant expected the transparency for the VR authentication protocol to be \textit{``well-defined and open-sourced''} for properly evaluating its security (P3). 

Another reason behind this gap is that participants prioritize interaction experience over security, resulting in an incorrect perception of the security of the authentication method. Figure~\ref{fig:overall_perception} and~\ref{fig:payment_pref}  show participants selecting what they considered the less secure, yet the more usable, \tap as their preferred authentication method. Participants attempted to bridge this gap by envisioning new alternatives of interaction modalities for all different concepts of authentication, such as new ways to interact with a card and drawing 3D signature \textit{``instead of a flat 2D traditional signature''} (P5). Along the same lines, researchers and developers have been actively improving both the usability and security of VR authentication~\cite{mathis2021fast,watson-chi22, olade-icvars20,zhu2020blinkey,liebers-chi21}.

\subsection{Challenges in Being Aware of Threats Across VR and the Real World.} 
Participants enjoyed being present and immersed in the VR environment. Meanwhile, they desired awareness of security threats in VR and the physical world. For example, when immersed in VR, participants wanted the VR application \textit{``to reflect the surrounding environment''} both virtually and physically for them to notice bystanders (P21). Another instance is one participant's need to know about suspicious activities in VR by \textit{``sending me [them] an SMS/email/etc.)''} even when they are back in the physical world (P8). 

Unfortunately, offering such cross-contextual awareness with an enjoyable experience is still an open challenge, even when the users develop a proper understanding of VR authentication. There are several reasons. First, VR creates an immersion effect, and users have limited perceptual capacities~\cite{tseng-chi22}, which is more challenging compared to more conventional digital mediums, such as websites and mobile apps. Furthermore, it is challenging to interpret suspicious activities when users are present in VR. Because the VR threats can be made more imperceptible, e.g., invisible card-skimming attacks (Section~\ref{sec:RQ2-2}), and users miss the physical context to better understand bystander activities in VR.

\subsection{Difficulties of Assessing Security Risks in Playful VR.}

Many users have high expectations for VR technology and applications, seeking creative and enjoyable experiences \cite{olsson-puc13, tang-chi23}. In our study, we found that participants also desired a playful authentication experience and proposed various ideas to achieve it. These ideas focused on improving usability and implementation, such as the interface to manage multiple virtual cards where users could \textit{``select the card from a pop up UI''} (P20) and informing payment success with \textit{``more fancy effects''} (P24). Participants also anticipated emotional appeals (e.g., enjoyment and playfulness) from the authentication interfaces, especially in the game context. 
\begin{displayquote} 
\textit{``I want the payment process to be cool and make me feel good. some visual effects could help and make me happy to make the payment.''} (P24) 
\end{displayquote} 
  
However, it is important to consider the trade-off associated with a playful authentication experience. As discussed in Section~\ref{sec:RQ2-3}, gamifying the VR experience can reduce participants' sensitivity to security risks and feedback. This reduction in awareness may have significant consequences and pose risks in security-sensitive scenarios, such as high-risk payments. This concern aligns with previous research conducted in the domain of digital games \cite{chen-chb10}.

\noindent
\colorbox{beaublue}{\textbf{Takeaway-RQ3.}} 
While participants prioritized interaction experience, the security of authentication is important to them. Participants expected both usability and security, but their expectations appeared conflicting. Our findings highlight three tensions. First, we noticed a gap between the perceived and actual security in VR authentication as some other contexts~\cite{khan-soups15}, and the VR interactions further exacerbate the discrepancy. Second, participants liked the immersion of VR while expressing concerns about threats in both the virtual and physical worlds. Third, although participants envisioned playful experiences, such experiences can diminish their sensitivity to security risks.

\section{Discussion}
Our findings yield recommendations to guide the design of future VR authentication to calibrate users' security perceptions, enhance VR systems' awareness of threats, and provide flexible feedback on security risks.

\paragraph{Exploring virtualized metaphors for calibrating security perceptions.} 

As we discussed in Section~\ref{sec:RQ2-1}, users have a limited and varying understanding of VR's inherent security properties, and how authentication interaction is virtualized may lead to misalignment between users' perceptions and the actual security properties. Calibrating such perceptions is challenging in the two following aspects, for which we provide corresponding suggestions. In general, we can actively leverage interaction metaphors~\cite{raja-soups11, angus-sdvrs95} in helping VR users align their security perceptions in a usable way. 

First, the security model of VR authentication can be more involved than the traditional platforms, including the use of unconventional and multi-modal modalities, e.g., eye tracking input and biometrics~\cite{mathis2021fast, liebers-chi21}. As such, the designers of VR authentication services could convey the security properties by \textit{ \textbf{integrating metaphors apt to VR authentication modalities}}. For example, when using eye tracking for two-factor authentication (PIN and biometrics), symbolizing eye interaction may help inform users of the use of biometrics when comfortably engaging them~\cite{lee-chi22}.   

On the other hand, the designer could \textit{ \textbf{better connect the context with authentication metaphors}}. We observed the positive effect that \tap reminded some participants of the payment context (Section~\ref{sec:RQ2-3}). We can further strengthen this association with the context by naturally embedding metaphors in the VR interactions with the primary task~\cite{micallef-soups17}. For example, to indicate enhanced security, one VR game can award users persistent credits after users make an effort to complete the multi-factor authentication. Another opportunity is to communicate security by storytelling in VR~\cite{kumar-chi18}.

\paragraph{System support for enhancing awareness of threats across VR and the real world}
As we found in Section~\ref{sec:RQ2-2}, VR users lack awareness and face difficulties in interpreting threats in both VR and the real world. These difficulties further prevent users from adequately assessing security. Though solutions exist to improve VR users' contextual awareness for safety reasons ~\cite{kudo-hci21}, it is not practical to rely on users only to comprehend all the security threats. Thus, we propose multiple improvements for the VR system to automatically detect threats across virtual and physical contexts and adapt security measures in authentication accordingly.  

First, we can design \textit{\textbf{intelligent systems to recognize and comprehend threats related to VR}}. Our observations in Section~\ref{sec:RQ2-2} and prior studies identified threats in users' virtual world (dark patterns~\cite{ruth-usenix19}, bystanders~\cite{stephenson-sp22}, etc.) and physical contexts (physical imposter~\cite{mathis2021fast}, unauthorized users~\cite{de-csur19}, etc.). The VR system may utilize machine learning to enhance their cross-contextual awareness~\cite{kudo-hci21}. It may comprehend the security implication of a virtual or physical entity based on the VR scene, physical environment, and multi-sensory inputs. For instance, the VR system can tell whether a bystander is actively observing the users during payment.  

Next, the VR system could employ \textit{\textbf{access control to safeguard users' virtual assets actively}}. Based on the contextual understanding, the VR platform could control other entities' access to one user's scene and asset, e.g., payment token, when they perform authentication. We noticed our participants' need for personalizing such access control for multi-user settings, e.g., to avoid \textit{"accident payment from the kids"} in a family with children (P20).  

Moreover, the VR system may \textit{\textbf{automatically adapt authentication methods according to threats in the context}} while balancing usability requirements. Though users prioritize usability over security, they still consider the security of authentication crucial (Section~\ref{sec:RQ3}). The VR system can adaptively enhance security measures to defend against perceived threats. For example, when detecting an active observer, the VR authentication may shuffle the PIN pad from the default. 

\paragraph{Multiplexing communication of security risk in gamified VR.} 
Current VR applications are predominately games, and users expect a playful experience (Section~\ref{sec:RQ3}). Section~\ref{sec:RQ2-3} explained how a gamified VR context might reduce people's sensitivity to risks and security feedback. Our findings in Section~\ref{sec:RQ3} echo prior work that users still demand transparency and control for payment authentication, e.g., properly understanding suspicious behavior. This complements the above recommendation to reduce users' effort in making security decisions. For security-sensitive applications, especially payment, risk communication is often necessary for users' assessment of security. Here we discuss the potential to improve the efficacy of risk communication by multiplexing the feedback in VR.   

VR applications deliver feedback over multiple modalities (e.g., visual, audio, and haptic) to their users. The payment authentication service may choose to \textbf{\textit{utilize orthogonal feedback modalities}} to the primary application context. As such, we avoid overwhelming users and causing security fatigue~\cite{harbach-chi14}. In addition, the application can interleave the timing of authentication feedback with the primary activities in VR. The choice of feedback modality also stands on an \textbf{\textit{understanding about VR users' perceptual capacity and sensitivity}}, which may also vary across different demographics, e.g., people with accessibility challenges~\cite{stephenson-sp22}.   

\paragraph{Future research direction.} Our work opens up several directions for future research, which stem from our findings, methodology, and limitations. 
First, our recommendations primarily focus on improving the perceived and actual security of VR authentication through effective security communication and infrastructural support. Furthermore, the VR ecosystem involves multiple stakeholders, such as platform providers, application developers, authentication service providers, and banks, each with distinct roles and responsibilities in payment authentication. Coordinating with these stakeholders to systematically enhance security poses an ongoing challenge.

Building upon the recommendations, future research could explore effective ways to assist users in making informed security decisions and implementing security measures, for example, making trade-offs between security and usability of authentication. More specifically, it would be valuable to investigate the optimal level of control to give users versus automating decision-making on their behalf~\cite{filipczuk-AAMAS22}.

In our user study's population, the participants primarily consisted of young and tech-savvy individuals who are often early adopters of VR technology. To generalize our findings, it would be valuable for researchers to explore how users from diverse backgrounds, including different age groups and levels of technology expertise, perceive VR authentication~\cite{ratakonda-idc19}.

In addition, future work may explore more possibilities of VR interaction for authentication with users involved in the co-design process~\cite{yao-chi19}. Through participatory design studies, experts and users can collaborate to design both the frontend interaction and backend infrastructure of VR authentication systems.

Next, we recommend conducting longitudinal research, such as diary studies~\cite{hayashi-chi11}, to examine users' long-term adoption and usage of VR authentication. Users' security attitudes and behaviors may evolve over time as they interact with the system~\cite{li-sp23}. These longitudinal studies will provide valuable insights into usability and security issues in real-world scenarios, including how users respond to suspicious activities and threats~\cite{downs-apwg07}.

\section{Conclusion}

We presented a technology probe study to investigate how interaction experiences in VR authentication affect users’ security perception. We designed four probes, using variants of PIN authentication, a virtual card, and a signature, that represent the paradigms of user authentication. We embedded these probes in the routine payments of a VR archery game. In our user study, we collected participants’ responses using surveys regarding interaction experiences, security perceptions, and expectations for authentication. We revealed how participants benefited from the virtualization of authentication in VR and faced unique challenges in interactions, e.g., motion control. Participants encountered difficulties and ambiguity due to VR interactions when transferring their prior authentication knowledge to the VR context. Participants' expectations centered around improving interaction factors with security remained a crucial but secondary factor. However, their expectations were conflicting. We identified tensions in their expectations, which drive our recommendations for future work.

\bibliographystyle{ACM-Reference-format}
\bibliography{main.bbl}
\end{document}